\documentclass[12pt]{iopart}

\usepackage{graphicx}
\usepackage{color}
\usepackage{amssymb}
\usepackage{algorithmic}
\usepackage[boxed]{algorithm}

\begin{document}

\title{Coexistence of extended and localized states in one-dimensional systems}

\author{A. M. C. Souza$^{1,2,4}$, R. F. S. Andrade$^{3,4}$ }

\address{$^{1}$Departamento de F\'{i}sica, Universidade Federal de
Sergipe 49.100-000, S\~{a}o Cristov\~{a}o - Brazil. \\

  $^{2}$Department of Physics, University of Central Florida, Orlando,
FL 32816, USA.\\

  $^{3}$ Instituto de F\'{i}sica, Universidade Federal da Bahia, 40210-340
Salvador Brazil.\\

  $^{4}$ National Institute of Science and Technology for Complex Systems.}

\ead{randrade@ufba.br}

\begin{abstract}
Mobility edge transitions from localized to extended states have been observed in two and three dimensional systems, for which sound theoretical explanations have also been derived. One-dimensional lattice models have failed to predict their emergence, offering no clues on how to actually probe this phenomenon in lower dimensions. This work reports results for a class of tight-binding models with electron-mass position dependence, for which localized-extended wave function transitions can be identified. We show that it is possible to control the density of localized and extended states by tuning the transition-related parameter for a continuous range of energy values. Mathematically exact results for extended or localized states are derived in two extreme conditions of this parameter, as well as an exact energy value for the mobility edge transition in the intermediate regime. Our framework provides a clear point of view on the phenomena and can also be harnessed for setting up experiments to probe to precisely evaluate the associated mobility edges using state-of-the-art technology.
\end{abstract}

\maketitle

\section{Introduction}

The localized-extended wave function transition in quantum systems
has been one of the most active areas of study in the condensed
matter physics during the last decades. As a matter of fact,
accurate phase transition diagrams based on the energy of the
quantum particle and a suitable control parameter explain most of
the experimental findings. In general, models based on the
one-electron tight-binding Hamiltonian are able to clearly identify
the transition the two types of wave functions. Among the different
systems, with one (or more) control parameters, that induce
localized-extended transition one may quote those displaying random
disorder \cite{anderson1958}, incommensurate potentials
\cite{aubry80}, quantum percolation \cite{chang1995,gong2009},
correlated disorder \cite{lyra}, and the variation of clustering
properties in complex networks \cite{jahnke2008,giraud2009}.

For systems that undergo mobility edge transitions, localized and
extended states can coexist at different energies: for a fixed value
of the control parameter, a critical energy level separates
localized from extended states, allowing for a fine tuning of the
systems conduction properties by simply altering the energy flux to
the system. Such a property adds substantial versatility to devices
and circuit components, what explains the large efforts made to
tackle this yet unsolved problem. We remind that the precise
experimental measurements of the mobility edge energy, which has
been actively pursued, was until recently an elusive issue even for
three dimensional systems\cite{modu15}. Similar subtleties are also
found in one-dimensional systems where, despite significant efforts,
further details on the presence of localized-delocalized transition
with mobility edges are still to be found on both theoretical and
practical sides \cite{ribeiro13,pasek17}. Indeed, the first
experimental evidences of such an effect in a one-dimensional
quasi-periodic optical lattice have been reported only recently
\cite{luschen17}. Those results are supported by a theoretical
description of the system in the continuum limit\cite{xli17}.
Surprisingly, after performing performing the tight-binding
approximation, it was found that the mobility edge disappears
\cite{xli17}. Indeed, we are lead to the well-studied Aubry-Andre
model, with a usual extended/localized transition where all states
are either localized or delocalized, depending only of the on the
strength of the control parameter \cite{aubry80,modu09}.

This work intends to fill that gap by analyzing the properties of a
well defined physical system on a regular lattice, where the
transition between extended and localized wave functions is clearly
gauged by one parameter associated with the effective electrons
mass. In particular, we consider a one-dimensional tight binding
model where the electron-mass is defined according to its position
along the lattice and reproduce the usual two- and three-dimensional
mobility-edge landscape, accounting for the presence of extended and
localized states in distinct regions of the spectrum. \textcolor[rgb]{0.00,0.00,0.00}{One amazing aspect of the model is the fact that it allows for exact analytical results in the two limiting cases, where only extended or localized states are present. These results are in full agreement with the numerical integration of the systems equations which, as usual in such circumstance, are also used to explore the intermediary regime where states of both classes coexist.}

\textcolor[rgb]{0.00,0.00,0.00}{The rest of this work is organized as follows: In Sec. II we present a brief review of the used formalism leading to position dependent masses, which is followed by the derivation of the equivalent tight-binding Hamiltonian in Sec. III.  Sec. IV is details the exact analytical treatment to account for the limiting behavior of the model, where only extended or localized states exist.  Results for the intermediate parameter value, where the two kinds of states coexist are discussed in Sec. V, while concluding remarks and possible applications of the formalism are brought in Sec. VI.
}
\section{Position dependent mass }

Models featuring position-dependent masses have been discussed in a
wide range of areas, from semiconductors
\cite{harr61,gora69,roos83,gell93} to quantum wells and quantum dots
\cite{harr00,kesh13,barba14,barba15}, and from astrophysics
\cite{rich82} to polarons \cite{zhao03} and nonlinear optics
\cite{khor12}. One way to introduce the effect of dependence of the
mass on the particle's position on the system's properties is to use
the framework of the position-dependent effective mass
Schr\"{o}dinger equations \cite{costa11,mazh12,costa13}. \textcolor[rgb]{0.00,0.00,0.00}{In such cases, position dependent masses are a consequence of imperfect translational invariance, caused mainly by lattice distortions and impurities. Indeed, if they are taken into account in a adequate way, their net effect can be included in the particles mass.} Such an
approach, which we follow in this work, has the advantage of
allowing to formally derive the corresponding tight-binding
Hamiltonian, which is used as starting point for all following
calculations. Once this Hamiltonian formalism is of widespread use
for the investigation of many quantum systems, the derivation we
present in the sequence may serve as a starting point for further
studies. We emphasize that Ref. \cite{costa13} presents an very amazing explanation on how the Morse potential, which has a widespread use in condensed matter physics, can be mathematically derived from the assumption of the position dependent mass.

Following Ref. \cite{costa11}, we start with the Schr\"{o}dinger
equation for a field-free particle in continuous space
\begin{equation} \label{eq1}
\hat{H}= - \frac{\hbar^{2}}{2m} \hat{D}_{\gamma}^{2} \psi (x),
\end{equation}
in which
\begin{equation} \label{eq2}
\hat{D}_{\gamma} = (1+\gamma x) \frac{d}{dx}
\end{equation}
shall be regarded as a deformed derivative translation operator,
where $\gamma\ge 0$ is a parameter associated with the
position-dependent effective mass $m_e$ of the particle.

The explicit relationship between $m_e$ and $\gamma$ is derived by
using Eq. (\ref{eq1}) to evaluate the system's average energy as

\begin{eqnarray} \label{eq3}
E &=& \langle H \rangle = - \frac{\hbar^{2}}{8} \int dx \left\{ \frac{d}{dx} \left[ \frac{d}{dx} \left( \frac{1}{m_e} \right) \right]  \phi \frac{d \psi }{dx} \right. \nonumber \\ & & \left. + \frac{2}{m_e} \phi \frac{d^{2} \psi}{dx^{2}}  +  \frac{d}{dx} \left[ \frac{d}{dx} \left( \frac{1}{m_e} \right) \right] \phi^{*} \frac{d \psi^{*} }{dx} \right. \nonumber \\ & & \left. +  \frac{2}{m_e} \phi^{*} \frac{d^{2} \psi^{*}}{dx^{2}} \right\} \;.
\end{eqnarray}

The expression $m_e \equiv m/(1+\gamma x)^{2}$ is obtained by
identification of the terms such that, if $\gamma\equiv0$, the
energy reduces to the usual expression for a particle with constant
mass. Under the same condition, the conjugate field of $\psi$,
expressed by $\phi=\frac{\psi^{*}}{(1+\gamma x)}$
\cite{rego13,rego16}, reduces to its usual form.

\section{The equivalent tight-binding Hamiltonian}

In order to move from the continuous formulation in Eqs.
(\ref{eq1}-\ref{eq3}) to a discrete one-dimensional space, we
consider that $x$ takes $N$ integer values $x_i$, where $N$ is the
number of allowed sites in a lattice. \textcolor[rgb]{0.00,0.00,0.00}{This is a crucial step in this work. Indeed, it represents an extension from the local position dependent mass, usually limited to vary within a unit cell or within a fixed width potential, to a general system with translational invariance. Nevertheless, the  results we obtain in the further development depend on the product $\gamma N$, much as observed for the fixed width  infinite potential \cite{costa11}. From the physical point of view, the mass dependency on position may then be thought as arising from a global distortion of the lattice, as a external field or continuously varying inertial effect in an optical assembled device \cite{luschen17}.}

Here, it is convenient to look
for the eigenstates in terms of the basis $\psi_{i} (x_{j}) =
\delta_{i,j}$ and $\phi_{i} (x_{j}) = \delta_{i,j} / (1 + \gamma
x_{j})$, where $\delta_{i,j}$ is the Kronecker delta. Expressing the
spatial derivatives in terms of differences between the pertinent
functions in two neighboring sites, the following matrix
representation of the system  Hamiltonian is obtained
\begin{eqnarray} \label{eq4}
H_{i,j} &=& - \frac{\hbar^{2}}{2ma^{2}} \{ (1+\gamma x_{j+1})
\delta_{i,j+1} + (1+\gamma x_{j-1})\delta_{i,j-1} \nonumber \\
& & -2 (1+\gamma
x_{j})\delta_{i,j} + \frac{\gamma a}{2} [ \delta_{i,j-1} -
\delta_{i,j+1} ] \}\, .
\end{eqnarray}
Here, $a$ is the discrete natural length associated with the lattice
spacing. We define $t\equiv \hbar^{2}/(2ma^{2})$ so that, from now
on, all lengths and energies are measured, respectively, in units of
$a$ and $t$.  We set $a=1$ and let the coordinates along the $x$
axis direction take the values $x_{j}=j$, with $j=1, 2, 3,.., N$, in
such a way that $\gamma x \ge 0$. Further, by setting $t=1$ and
representing the annihilation and creation fermionic operations by
$\hat{c_{j}}$ and $\hat{c}_{j}^{\dag}$, we can write the nearest
neighbor tight-binding Hamiltonian as
\begin{equation} \label{tb}
\hat{H}= \hat{H}_{0} + \hat{H}_{1} \;,
\end{equation}
where
\begin{equation} \label{tb0}
\hat{H}_{0}= - \sum_{j} (\hat{c}_{j}^{\dag} \hat{c}_{j+1} +
\hat{c}_{j+1}^{\dag} \hat{c}_{j} ) \;,
\end{equation}
and
\begin{equation} \label{tb1}
\hat{H}_{1}= \gamma \sum_{j} \left\{ 2j \hat{c}_{j}^{\dag}
\hat{c}_{j} + \frac{(2j+1)}{2}  (\hat{c}_{j}^{\dag} \hat{c}_{j+1} +
\hat{c}_{j+1}^{\dag} \hat{c}_{j} ) \right\} \; .
\end{equation}

The energy spectrum and the wave functions of the Hamiltonian (\ref{tb})
are obtained through $\hat{H} |\psi \rangle = E |\psi \rangle$,
where the eigenstates of $\hat{H}$ corresponding to eigenvalues $E$
can be expanded as $|\psi \rangle = \sum_{j} b_{j} |j \rangle
=\sum_{j} b_{j} \hat{c}_{j}^{\dag} |\emptyset \rangle $, with
$|\emptyset \rangle$ denoting the vacuum state. From Eqs.
(\ref{tb}), (\ref{tb0}) and (\ref{tb1}), we obtain the following
recurrence equation for the coefficients $b_{j}$
\begin{eqnarray} \label{exp2}
& &\left( \gamma(j+\frac{1}{2}) + 1 \right) b_{j+1} + \left( \gamma(j-\frac{1}{2}) + 1 \right) b_{j-1} \nonumber \\ & & = \left( 2 \gamma j - E \right) b_{j} \; ,
\end{eqnarray}
which must satisfy the boundary conditions $b_{0}=b_{N+1}=0$.  We would like to call the attention that the dependency of the hopping terms on the position j appears in a similar equation derived to describe the behavior of  periodically driven systems \cite{jensen89,oliveira94}.

We observe that the problem of characterizing the spectra of
$\hat{H}_{0}$ and $\hat{H}_{1}$ can be carried out exactly for each
one of the operators. However, as they do not commute, the overall
Hamiltonian $\hat{H}$ could not be solved in the same way. Hence,
before addressing the general problem, let us briefly take care of
the eigenstates of $\hat{H}_{0}$ and $\hat{H}_{1}$.

\section{Analytical results for limiting conditions}

$\hat{H}_{0}$ is the usual tight binding Hamiltonian for the
homogeneous one-dimensional lattice. When $\gamma=0$, Eq.
(\ref{exp2}) reduces to  $( b^{(0)}_{j+1} + b^{(0)}_{j-1} ) = -E_{0}
b^{(0)}_{j}$, which coincides with the  recurrence relations of the
$U_{j}$ Chebyshev polynomials of the second kind of degree $j$. That
leads to the eigenvalues $E_{0}(j)=-2 \cos \left( \frac{j\pi}{N+1}
\right)$ and wave function coefficients
$b^{(0)}_{j}=\frac{1}{\sqrt{N+1}} \sin \left(\frac{j\pi}{N+1}
\right)$. This well known result characterizes Bloch extended states
for all energies in the spectrum, with density of states (DOS)
\begin{equation} \label{br}
\rho_{0} (E) = \frac{1}{\pi \sqrt{4-E^{2}}}, \;\;\;\;\; E
 \in [-2,2] \;
\end{equation}
in the $N \rightarrow \infty$ limit.

As for the eigenvalues and eigenvectors of $\hat{H}_{1}$, we
consider only the terms that are multiplied by $\gamma$ in Eq.
(\ref{tb1}), which leads to
\begin{equation} \label{exp3}
\left( j+ \frac{1}{2} \right) b^{(1)}_{j+1} + \left( j- \frac{1}{2}
\right) b^{(1)}_{j-1} = 2 \left( j - \epsilon \right) b^{(1)}_{j}
\;,
\end{equation}
where $\epsilon$ indicates the scaled energy, i.e., $\epsilon =
\frac{E}{\gamma}$. In the $N \gg 1$ limit and for sufficiently large
$j$, Eq. (\ref{exp3}) can be approximated by $(j+1) b^{(1)}_{j+1} +
(j-1) b^{(1)}_{j-1} \cong 2 \left( j - \epsilon \right)
b^{(1)}_{j}$, which is the recurrence relation of the Laguerre
polynomials $L^{(-1)}_{j}$. Using the same boundary conditions as
above and taking into account the proper asymptotic expansions
\cite{szego39}, we can write the eigenvalues as
\begin{equation} \label{eig1}
E_{1} (j)=4 \gamma (N+1)\cos^{2}(\theta_{j}),
\end{equation}
where $\theta_{j} \in (0,\pi/2)$ is obtained by the $N$ equations
\begin{equation} \label{eig2}
(N+1)[\sin(2\theta_{j})-2\theta_{j}] + \frac{3\pi}{4} = j\pi \;,
\;\;j=1,2,...,N.
\end{equation}
After straightforward calculations, it is possible to show that the
DOS in the $N \rightarrow \infty$ limit is given by
\begin{equation} \label{eig3}
\rho_{1} (E) = \frac{\sqrt{E(4\gamma N-E)}}{ 2\gamma \pi N E }
\;\;\;\;\; E/\gamma N \in  [0,4] \;.
\end{equation}

Therefore, the eigenstates of $\hat{H}_{1}$, characterized by
Laguerre polynomial coefficients, feature localized-like behavior
over the whole energy spectrum. As a matter of fact, the asymptotic
behavior $b^{(1)}_{j} \sim j^{-3/4}$\cite{szego39} shows that the
wave functions decay polynomially. In seemingly contrast with the
spectrum of extend states, Eq. (\ref{eig3}) indicates that the
allowed energy interval increases linearly with the size of the
system for fixed $\gamma$. However, to analyze the results in the
limit of large systems it is more convenient to assume the explicit
dependency $\gamma N = constant$.

\begin{figure}[h!]
\centering
\includegraphics[width=8cm,angle=0]{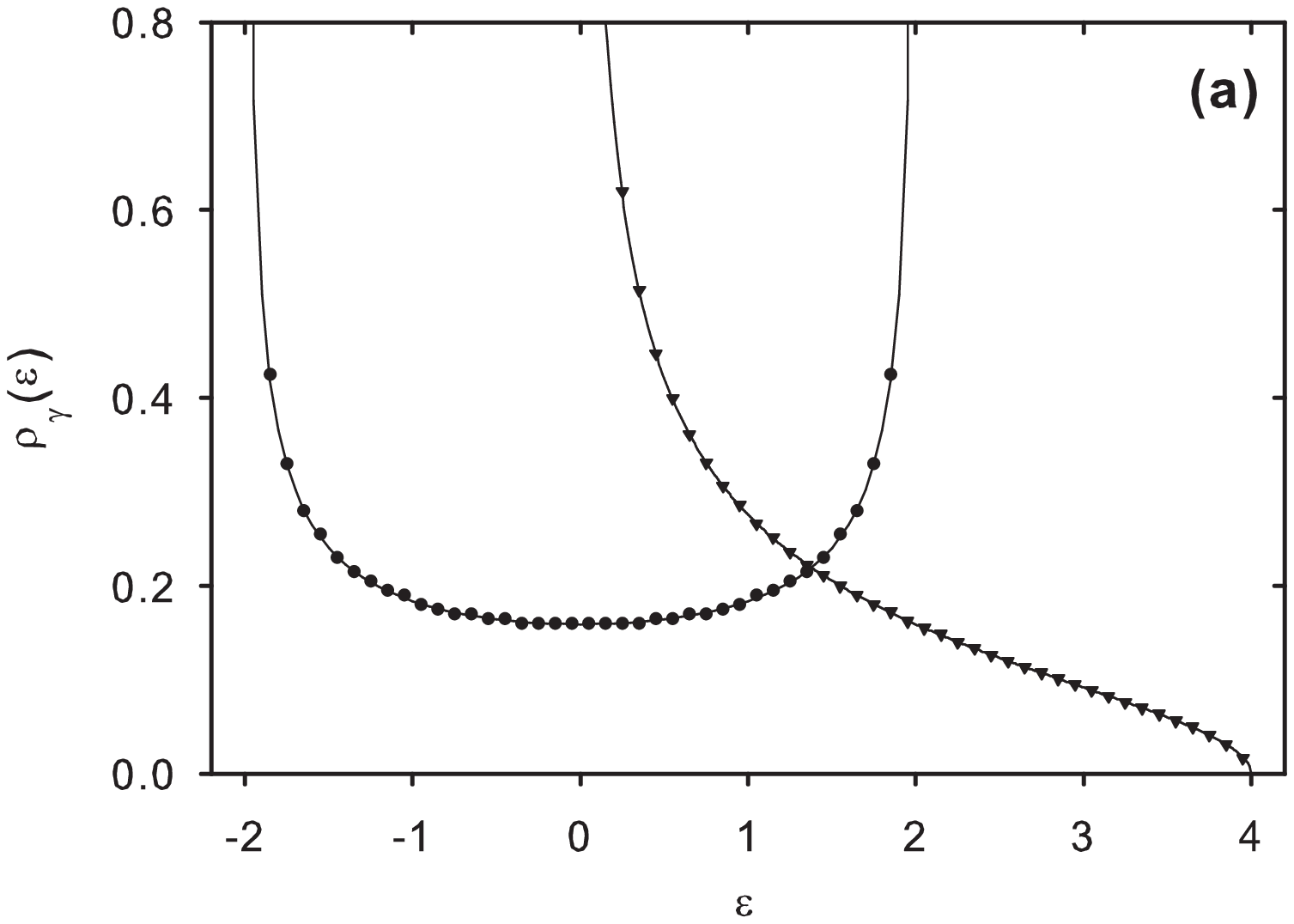}
\includegraphics[width=8cm,angle=0]{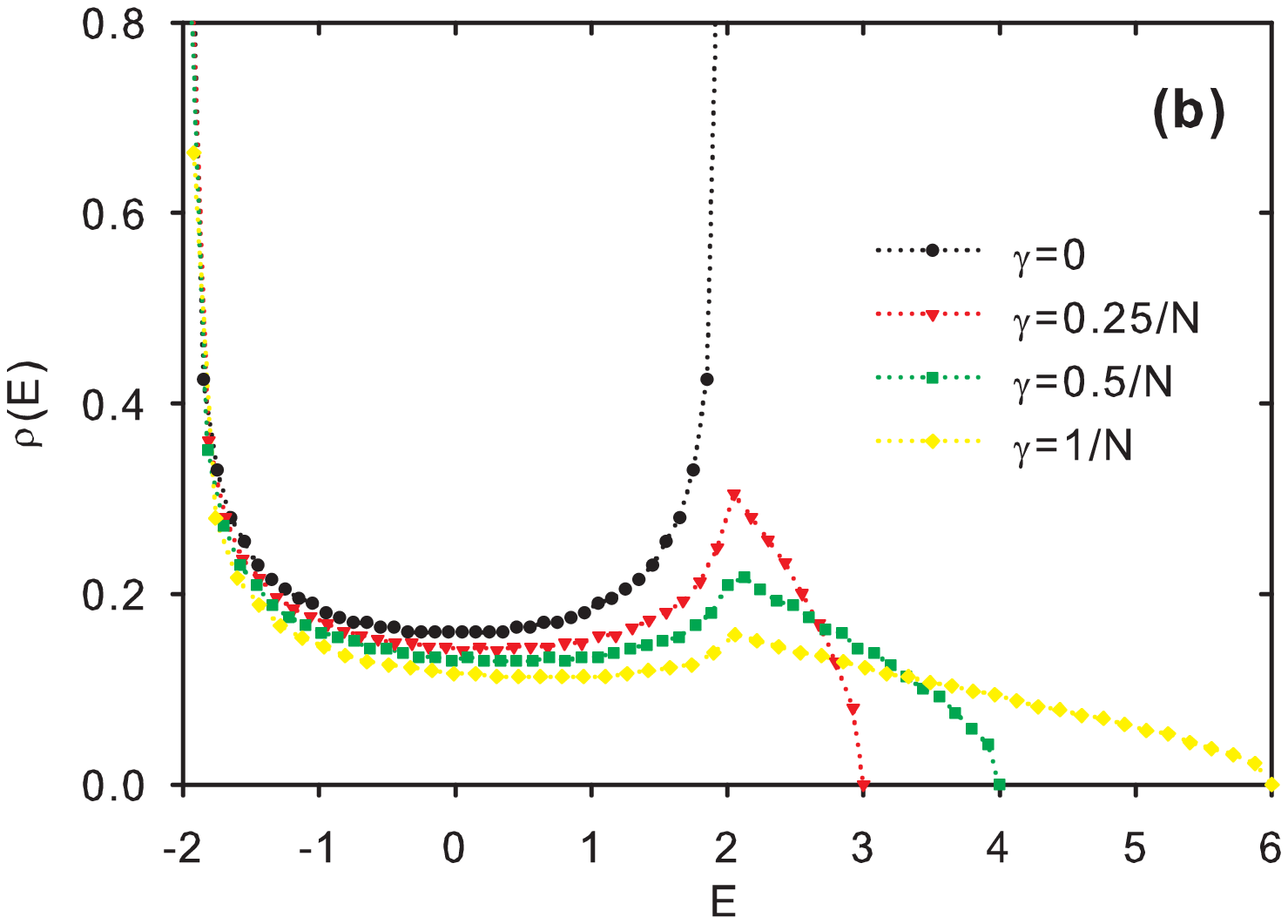}
\caption{ (a) DOS $\rho (E)$ as a function of $E$ for $\hat{H}_{0}$
(circles) and $\hat{H}_{1}$ (triangles) for $\gamma = 1/N$ from a
system of $N=8000$ sites. Solid lines correspond to the analytical
expressions in  Eqs. (\ref{br}) and (\ref{eig3}). (b) DOS for
$\hat{H}$ as a function of $E$ for some typical values of $\gamma$
in the \emph{intermediate regime}. } \label{ceroa}
\end{figure}

Fig. \ref{ceroa}(a) shows the DOS for $\hat{H}_{0}$ and
$\hat{H}_{1}$ given by, respectively, Eqs. (\ref{br}) and
(\ref{eig3}). We also draw the energy density of states obtained by
numerical diagonalization of the two corresponding  system of
$N=8000$ sites. It is clearly shown that the approximate expression
for the  $\hat{H}_{1}$ (Eq. (\ref{eig3})) agrees with the numerical
results, in such a way that, in this limit, the approximate and
numerical results coalesce with the exact result for $\rho_{1} (E)$.

The individual solutions for $\hat{H}_{0}$ and $\hat{H}_{1}$ allows
the identification of two $\gamma$ dependent limiting conditions for
$\hat{H}$. When $\hat{H}_{0}$ plays the relevant role, every energy
level corresponds to a Bloch extended wave function. On the other
hand, when $\hat{H}_{1}$ becomes relevant, every energy level has a
Laguerre localized wave function. Therefore, the extended/localized
regimes are controlled by the space dependent effective mass through the
parameter $\gamma$.

\section{Results for the coexistence regime}

Differently from the cases when $\hat{H}_{0}$ and $\hat{H}_{1}$
control the system separately, we cannot find a general analytical
solution for $\hat{H}$. While exact solutions in the particular
limits reproduce the results for $\hat{H}_{0}$ and $\hat{H}_{1}$,
the general case requires analytical approximation approaches and
the numerical method of cluster exact diagonalization. Fig.
\ref{ceroa}(b) illustrates the behavior of $\rho(E)$ for selected
choices of $\gamma$ in $\hat{H}$ obtained numerically for a $N=8000$
sites system.  The numerical results consistently indicate that the
energy band extends itself from $E=E_{min}=-2t$ to
$E=E_{max}=2t+4\gamma N$, values that correspond to the sum of lower
and upper bounds from the two individual models shown in Fig.
\ref{ceroa}(a).  For any finite $\gamma N$, the energy band is
finite. In accordance to the above relation, when $\gamma=1/N^2$ the
result for $\rho_0(E)$ is recovered. More generally, our numerical
investigations based on a general relation $\gamma \sim N^{-\alpha}$
suggest that, in the $N \rightarrow\infty$ limit, the upper bound
converges to $2t$ if $\alpha>1$, whereas it diverges when
$\alpha<1$. Thus, the analytical results obtained from the analysis
of Eq. (\ref{exp3}) prevails for the full Hamiltonian $\hat{H}$.

Three different regimes can be devised for the model:
\begin{enumerate}
  \item Bloch regime: $0 \leq \gamma < O( N^{-1})$
  \item Intermediate regime: $\gamma \sim O( N^{-1})$
  \item Localized regime: $\gamma > O( N^{-1})$
\end{enumerate}

In the \emph{Bloch regime}, it is easy to see that Eq. (\ref{exp2})
reduces to the same form as for $\gamma=0$. Thus, only $\hat{H}_{0}$
plays a relevant role and every energy level has a Bloch extended
wave function. On the other hand, in the \emph{localized regime} where only $\hat{H}_{1}$ is relevant, every energy level has a Laguerre
localized wave function and the energy spectrum is not limited.
Despite an intensive search in the literature, we were not able to find any investigation on an actual physical system where the range of values of $\gamma$ satisfy this condition.

Let us then consider the \emph{intermediate regime} as it does not only comprise our most important findings, but also allows for establishing a direct relationship to a real world application. In the case of Ge quantum dots, experimental measurements show that a interface potential is responsible for quantum confinement, the extension of which can be traced back to the value of $\gamma$ \cite{barba14,barba15}. Amazingly, within the carrier effective-mass formalism, the functional dependence between $\gamma$ and the size of this region is exactly the same as that of the intermediate regime defined above.

In order to make use of analytical asymptotic expressions, let us assume the non-binding restriction $N \gg 1$. For a fixed $\gamma N$, we
identify the coexistence of extended and localized states. The
$\hat{H}$ eigenfunctions entail contributions from the $\hat{H}_{0}$
extended states as well as from the $\hat{H}_{1}$ localized states.
However, depending on the energy eigenvalue, the eigenfunctions
display more localized or extended properties. To quantify this
contribution, we numerically investigated the local properties of
the wave function by evaluating the participation ratio
\begin{equation}\label{PR}
  \zeta(E) = \frac{1}{\sum_{j} |b_{j}(E)|^{4}}.
\end{equation}
The localized or extended character of the states can be inferred
from the dependency of the value of $\zeta(E)$ with respect to the
system size $N$: it decays to zero for localized states or converges
to a finite non-zero value for extended states.

\begin{figure}
\centering
\includegraphics[width=8cm,angle=0]{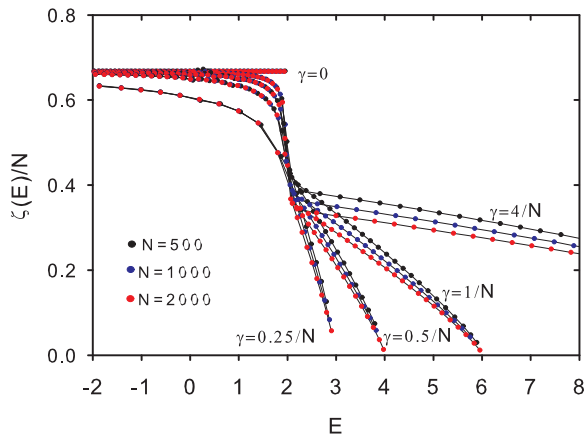}
 \caption{Dependence of $\zeta(E)$ as a function of $E$ and $N$ for a selected number of states in the allowed energy range. The mobility edge separating extended and localized states occurs always at $E=2, \, \forall \gamma N$ values.} \label{ceipr}
\end{figure}

Figure \ref{ceipr} shows the participation ratio $\zeta_{E}$ as a
function of the energy $E$ for $N=500$, $1000$ and $2000$ and
$\gamma=0$, $0.25/N$, $0.5/N$, $1/N$ and $4/N$. Our results show
that, on increasing the value of $N$, $\zeta_{E}$ has a clear
monotonic decay to $0$ when $E>2$, while it remains at a finite
value when $E<2$. Thus, the mobility edge occurs always at the value
$E=2$, irrespective of the product $\gamma N$.

\begin{figure}
\centering
\includegraphics[width=8cm,angle=0]{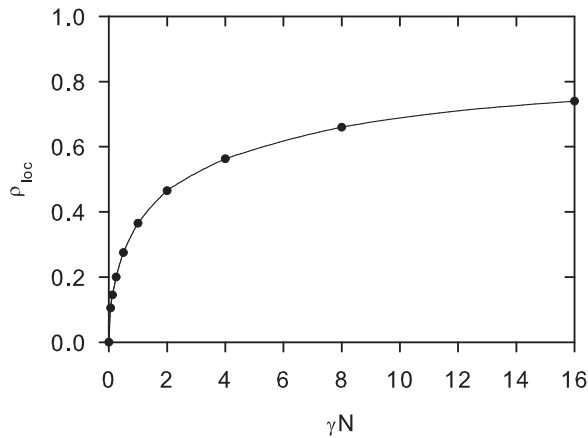}
\caption{Behavior of $\rho_{loc}(\gamma)$ as a function of $\gamma$ in the $N \rightarrow \infty$ limit. The numerical results indicate that a nonzero fraction of localized states is observed as soon as $\gamma >0$.} \label{ceden}
\end{figure}

A further important measure to compute is the fraction of energy
levels corresponding to localized and extended states and to obtain
its dependency on $\gamma$. In Figure \ref{ceden} we show the
density of localized levels $\rho_{loc}(\gamma)$ as a function of
$\gamma$ in the $N \rightarrow \infty$ limit. All of such states are
restricted to $E > 2$, which corresponds to the region of
localized states. The results are estimated from numerical
evaluations for $N=200$, $400$, $800$ and $1600$.  The numerical
results indicate that the fraction of localized levels in the total
spectrum increases with $\gamma$, but the fraction of extended
states never vanishes. It is interesting to note that, in the case
of free electrons in the presence of an uniform strong electric
field, localized and extended states have also been found. However,
the extended states are restricted to the edge of the band and the
fraction of these levels in the total spectrum goes to zero at the
large $N$ limit\cite{elet73}.

\section{Conclusions}
Our results clearly indicate the coexistence of extended and localized quantum states in a simple one-dimensional system of particles with position dependent masses. Although our work goes along the same direction as recent efforts to characterize theoretical aspects of the problem \cite{jahnke2008,pasek17,luschen17,xli17}, its very simple form allowing for exact results does provide another point of view on the subject, which might be useful to new experimental designs \cite{modu15}.

The presence of a position dependent mass $m_e$, leading to the coupling between the momentum and position operators in the Schr\"{o}dinger equation as well as in the $\hat{H}$ Hamiltonian, is the key feature of our approach. Although the presence of linear terms in a Schr\"{o}dinger equation is a common feature whenever the systems stays under the influence of electric field, the particular functional dependence caused by $m_e$ is much richer and induced the features discussed herein. The numerical results can be traced back to well documented results that we obtained analytically for each limiting regime, in which either localized or extended states are to be found. That ultimately led us to set a well-defined energy value for the mobility edge. Besides showing the direct relationship of our results to the confinement potential in a carrier space dependent effective mass scenario, we conjecture whether the recent experimental advances indicating mobility edge transition in one-dimensional optical lattice might benefit from the robust theoretical framework reported in this work.

\section{Acknowledgements}

We thank G.M.A. Almeida for valuable suggestions on the manuscript
and the financial support of Brazilian agency CNPq. Both authors
benefit from the support of the Instituto Nacional de Ci\^{e}ncia e
Tecnologia para Sistemas Complexos (INCT-SC).

\end{document}